\documentclass[fleqn,10pt]{wlscirep}
\usepackage[utf8]{inputenc}
\usepackage[T1]{fontenc}
\usepackage{mathcomp}

\usepackage{bm}
\usepackage{amsmath}

\title{Importance of \textit{\textbf{d}}$_{\textit{\textbf{xy}}}$ orbital and electron correlation in iron-based superconductors revealed by phase diagram for 1111-system}

\author[1]{Tsuyoshi Kawashima}
\author[1,*]{Shigeki Miyasaka}
\author[1]{Hirokazu Tsuji}
\author[1]{Takahiro Yamamoto}
\author[1]{Masahiro Uekubo}
\author[1]{Akira Takemori}
\author[1,2]{Kwing To Lai}
\author[1]{Setsuko Tajima}
\affil[1]{Department of Physics, Osaka University, Toyonaka, Osaka, 560-0043, Japan}
\affil[2]{Department of Physics, The Chinese University of Hong Kong, Shatin, Hong Kong}
\affil[*]{miyasaka@phys.sci.osaka-u.ac.jp}

\begin{abstract}

The structural flexibility at three substitution sites in LaFeAsO enabled investigation of the relation between superconductivity and structural parameters over a wide range of crystal compositions. 
Substitutions of Nd for La, Sb or P for As, and F or H for O were performed. 
All these substitutions modify the local structural parameters, while the F/H-substitution also changes band filling. 
It was found that the superconducting transition temperature $T_{\rm c}$ is strongly affected by the pnictogen height $h_{Pn}$ from the Fe-plane that controls the electron correlation strength and the size of the $d_{xy}$ hole Fermi surface (FS). 
With increasing $h_{Pn}$, weak coupling superconductivity switches to the strong coupling one where the $d_{xy}$ hole FS is crucially important.

\end{abstract}
\DeclareUnicodeCharacter{2212}{-}
\begin{document}

\flushbottom
\maketitle
%
%
\thispagestyle{empty}


\section*{Introduction}

Identifying the mechanism of high-temperature superconductivity has been a challenging task in solid state physics. 
For iron pnictide superconductors\cite{kamihara2006iron,kamihara2008iron}, more than ten years' effort has been devoted to clarifying the superconductivity mechanism. 
However, the essential problem of a pairing glue is still under debate. 
In many iron-based superconductors, the superconducting phase is observed near the quantum critical point of magnetic order\cite{jiang2009superconductivity} and/or nematic order\cite{kasahara2012electronic}. 
This suggests a spin or orbital fluctuation as pairing glue. 
An alternative proposal emphasizes strong electron correlation near the Mott insulator regime\cite{misawa2012ab,de2014selective,misawa2014superconductivity}, rather than a magnetic or orbital fluctuation in the weak-coupling regime.

Experimentally, it is well known that there is a correlation between the superconducting transition temperature ($T_{\rm c}$) and local crystal structure parameters, such as the pnictogen height ($h_{Pn}$) from the Fe-layer, and the As-Fe-As bond angle ($\alpha$). 
First, Lee $et$ $al.$ demonstrated that $T_{\rm c}$ reaches a maximum when $\alpha$ is close to 109.47$^\circ$, corresponding to the value for a regular FeAs$_4$ tetrahedron\cite{lee2008effect}. 
A similar correlation was found between $T_{\rm c}$ and $h_{Pn}$\cite{mizuguchi2010anion}. 
Here the maximum $T_{\rm c}$ was observed at $h_{Pn}\sim 1.38$ \AA \ . 
The main questions are: which physical parameters are controlled by these structural parameters, and how do they contribute to superconductivity?

According to theoretical calculations, the $d_{xy}$ hole Fermi surface (FS) around $(\pi, \pi)$ in the unfolded Brillouin zone expands with increasing $h_{Pn}$\cite{kuroki2009pnictogen}. 
However, it is not obvious how this electronic change controls superconductivity. 
In fact, in the LaFeAsO (La-1111) system, $T_{\rm c}$ does not vary monotonically with changes in lattice parameters\cite{miyasaka2013two,Lai2014evolution,miyasaka2017three}. 
Our previous studies have revealed that there are three superconducting phases in LaFe(As,P)(O,F/H), depending on the compositions\cite{miyasaka2017three}. 
Although bosonic fluctuation via FS nesting was discussed as the most plausible pairing mechanism, the high $T_{\rm c}$ in the third superconducting phase (SC3) could not be explained using the same scenario, because the nesting condition is very bad. 
Here it is unclear which structural parameter switches the mechanism, and how. 
As an alternative to the FS nesting based weak-coupling model, it is worth examining a strong-coupling scenario based on strong electron correlation near the $d^5$ Mott insulator regime\cite{misawa2012ab,misawa2014superconductivity}.
Here it is unclear which structural parameter switches the mechanism, and how.

In this study, we extended our previous work on LaFe(As,P)(O,F/H)\cite{miyasaka2017three} by covering a wider range of structural parameters through three site substitutions (Nd for La, Sb or P for As, and F or H for O). 
All these substitutions modify the local structural parameters, while the F/H-substitution also changes band filling. 
Precise measurements of structural parameters and resistivity for all samples revealed that the $d_{xy}$ hole FS is crucially important in a wide doping range for the high $h_{Pn}$ compounds. 
Although there are apparently three SC phases, two of them (SC1 and SC3) turn out to be of the same origin. 
It is likely that with increasing $h_{Pn}$ electron correlation becomes strong, which switches the FS nesting based weak coupling superconductivity (SC2) to strong coupling superconductivity derived from orbital-selective Mott systems (SC1/SC3). 
$T$-linear resistivity was commonly observed near the $T_{\rm c}$-maximum composition in each phase.

\section*{Results}

First, we present structural data for several typical compositions. 
(Data for all remaining compositions are provided in the Supplementary Information.) 
Figures 1(a), (b), (f) and (g) show the lattice constants $a$ and $c$ as functions of P/Sb content ($x$) for $R$FeAs$_{1-x}$(P/Sb)$_x$O$_{1-y}$(F/H)$_y$ with $y=0.1$ and $0.3$ ($R$ = La and Nd). 
Part of the dataset is taken from our previous papers\cite{miyasaka2013two,Lai2014evolution,miyasaka2017three}. 
The P/Sb contents ($x$) were determined by energy-dispersive X-ray spectroscopy (EDX), while the F/H compositions ($y$) are nominal values. 
Here we see that the lattice constants vary linearly with $x$ according to Vegard’s law, demonstrating the successful substitution of P or Sb for As. 
The result indicates that P-substitution has the effect of lattice compression, while the Sb-substitution has the opposite effect.
More detailed structural parameters, such as the pnictogen ($Pn$) height from the Fe-plane ($h_{Pn}$), the $Pn$-Fe-$Pn$ bond angle ($\alpha$) and the Fe-$Pn$ bond distance ($d_{{\rm Fe}-Pn}$) are presented in Figs. 1(c), (d), (e), (h), (i), and (j). 
These parameters also vary linearly with $x$. 
With increasing Sb-content, thereby expanding the lattice, $h_{Pn}$ and $d_{{\rm Fe}-Pn}$ increase, while $\alpha$ decreases. 
In Nd-systems, although the lattice constants are smaller, $h_{Pn}$ is larger and $\alpha$ is smaller than in the La-systems. 
As will be discussed later, among all these parameters, $h_{Pn}$ and/or $\alpha$ plays the most crucial role in determining the electronic state. 
From the present results, we find that the common primary effect of P/As(Sb), O/F(H) and La/Nd substitutions is to increase $h_{Pn}$.

To observe the change in carrier interaction, we measured the temperature ($T$) dependence of resistivity ($\rho$). 
In Fig. 2, the $T$-dependence of $\rho-\rho_0$ is plotted on a log-scale, where $\rho_0$ is residual resistivity. 
In all systems, with increasing P-content $x$, $\rho-\rho_0$ changed from non-Fermi liquid-like $T$-linear behavior to $T^2$-behavior. 
This implies that As-rich compositions are in the strong interaction regime, while with increasing P-content, the interaction is weakened, and the system becomes a Fermi liquid. 
Accordingly, as seen in the insets, $T_{\rm c}$ decreases with increasing P-content.

In Fig.3, we plot $T_{\rm c}$ and the power $n$ in $\rho (T)=\rho_0 + AT^n$ for all the P-doped samples in Fig. 2. 
Figures 3(a) and (b) show the results for $y=0.1$. 
In both La- and Nd-systems, at P100\% composition ($x=1.0$), $T_{\rm c}$ is very low (about 5 K) and $n\sim 2$. 
With decreasing $x$, $T_{\rm c}$ gradually increases, while the power $n$ monotonically decreases towards unity. 
This clear correlation between $T_{\rm c}$ and $n$ demonstrates that the key interaction for superconductivity is strengthened with decreasing $x$, as we have previously reported \cite{miyasaka2013two,Lai2014evolution,miyasaka2017three}. 
When $x$ decreases below $x=0.4$, $T_{\rm c}$ increases further in the Nd-system, while it is almost unchanged in the La-system. 
Our previous studies demonstrated that there is a critical change in the electronic state at around $x=0.4$. 
This was evidenced not only by transport properties such as resistivity and Hall coefficient, but also by direct observation of the band structural change through the angle-resolved photoemission experiment\cite{takemori2018change}. 
We distinguish these two superconducting regions, naming the As-rich region “SC1” and the P-rich region “SC2”.

According to theoretical calculations by Kuroki $et al.$\cite{kuroki2009pnictogen}, when $h_{Pn}$ increases, the $d_{xy}$ hole FS around $(\pi, \pi)$ becomes larger. 
Therefore, when $h_{Pn}$ decreases with $x$, as indicated in Fig. 1, the $d_{xy}$ hole FS is expected to shrink and eventually disappear as illustrated in Fig. 3(e). 
We believe this Fermi surface change drives the phase change from SC2 to SC1. 
The latter can be defined as the phase in which the $d_{xy}$ hole FS contributes to superconductivity, while this is not expected in SC2\cite{tin2020}.

For heavy electron-doping ($y=0.3$), the $n$-value is much larger than one at $x=1.0$, while it rapidly decreases when approaching $x=0$. (See Figs. 3(c) and (d).) 
At $x=0$ where the high $T_{\rm c}$ of over 40 K is measured, $n$ is close to 1, namely $T$-linear resistivity is observed. 
It is a common feature in the iron-based superconductors that while the P-rich compound is a weakly-interacting (Fermi-liquid like) system, the interaction becomes stronger with increasing As-content, which induces superconductivity and increases $T_{\rm c}$. 
A significant difference from the lightly-doped case ($y=0.1$) is that superconductivity disappears with P-substitution. 
As depicted in Fig. 3(f), we expect that, with heavy electron doping, the electron FS becomes much larger than the hole FS at the zone center. 
This imbalance of hole- and electron-FS must prevent FS nesting-derived superconductivity in the P-rich compositions. Therefore, superconductivity in the P-rich compounds can be understood within the FS nesting scenario over a wide doping ($y$) range. 

Since the $d_{xy}$ hole FS at the zone corners shrinks and eventually disappears with increasing P-doping level $x$\cite{usui2015origin}, we conclude that the $d_{xy}$ orbital is crucially important for superconductivity for heavily electron doping. 
In the Nd-system, $T_{\rm c}$ remains high up to a larger $x$-value than in the La-system. 
This can be attributed to the larger $d_{xy}$ hole FS due to the higher $h_{Pn}$ in the Nd-system compared to the La-system.

We distinguish this heavily electron-doped region from the lightly doped region (SC1) by labelling it SC3, because the $T_{\rm c}$ is strongly suppressed in the intermediate range of $y$ in the La-system. 
However, in the Nd-system, there is no such $T_{\rm c}$ suppression, and high $T_{\rm c}$ is maintained over a wide $y$-range. 
Considering that the difference between the La- and Nd-systems is caused by the greater pnictogen height ($h_{Pn}$), we enlarged $h_{Pn}$ by Sb substitution for As in the La-system. 
An interesting example at $y=0.14$ is shown in Fig. 4. 
In the Nd-system (Fig. 4(b)), the $x$-dependence of $T_{\rm c}$ at this doping level is qualitatively the same as that for $y=0.10$. 
However, in the La-system (Fig. 4(a)), $T_{\rm c}$ is strongly depressed near $x=0$, while it recovers with increasing Sb-content, probably because of the enlargement of the $d_{xy}$ FS via the increase in $h_{Pn}$. 
This suggests that SC1 and SC3 have the same origin.

\section*{Discussion}

Figures 5(a) and (b) are contour plots illustrating how $T_{\rm c}$ changes with $x$ and $y$ for the La- and Nd-systems, respectively. 
The dots represent the compositions studied in the present work. 
In these phase diagrams, we distinguish three superconducting phases (SC1, SC2 and SC3), particularly in the La-system. 
Compared to the change from SC1 to SC2 with the disappearance of the $d_{xy}$ FS, the change from SC1 to SC3 is not clearly defined. 
In the La-system, although these two phases are separated at $x=0$, when Sb is substituted, they merge with each other,  as in the Nd-system, owing to the enlargement of the $d_{xy}$ FS by the increase in $h_{Pn}$. 
We therefore think that the superconductivity mechanisms in phases SC1 and SC3 have the same origin. 
According to this scenario, the $T_{\rm c}$ suppression near $x=0$ for the intermediate doping range ($0.14<y<0.25$) is not caused by the bad FS nesting condition, but by the loss of $d_{xy}$ FS due to electron doping. 
The lost $d_{xy}$ FS can be recovered by Sb substitution and/or electron doping.

To visualize the relation between the local structural parameters and $T_{\rm c}$, Fig. 6 presents a contour plot of $T_{\rm c}$ in the $\alpha$-$d_{{\rm Fe}-Pn}$ plane. 
Here the blue broken lines represent constant $h_{Pn}$, and the star symbol indicates the point corresponding to a regular tetrahedron of Fe$Pn$. 
For each As/P(Sb)-substitution system, the SC phases lie on a straight line directed from lower right to upper left in the order SC2, SC1 and SC3. 
With electron doping, the bond angle ($\alpha$) decreases. 
Moreover, $T_{\rm c}$ increases in the order of SC2, SC1 and SC3 as the structural parameters approach those for the regular tetrahedral structure.

This figure indicates that the local crystal structure strongly affects the $T_{\rm c}$ value in Fe-based superconductors. 
In SC3 phase where $h_{Pn}$ is close to 1.38 \AA, $T_{\rm c}$ is very high and insensitive to the band filling. 
Note that band filling should be another important parameter in weak coupling superconductivity where bosonic fluctuation via FS nesting acts as a pairing glue. 
In fact, in Fig.6, the $T_{\rm c}$ suppression due to electron doping is visualized by the white bands extending from upper left to lower right. 
Therefore, in SC2 phase (P-rich compositions), the FS-nesting condition is crucially important, which supports a weak coupling regime\cite{mazin2008unconventional,kuroki2008unconventional,kontani2010orbital}. 
However, in the large $h_{Pn}$ region, there is no white area, namely, the FS-nesting condition does not affect $T_{\rm c}$ at all. 

An alternative scenario is strong electron correlation near the $d^5$ Mott insulator regime\cite{misawa2012ab,de2014selective,misawa2014superconductivity}. 
In this scenario, the total electron correlation is strong enough to induce antiferromagnetic order in LaFeAsO. 
Superconductivity is realized when the carriers are doped in this strongly-correlated parent compound. 
The electron correlation is enhanced with increasing $h_{Pn}$. 
Because $h_{Pn}$ can be controlled by As(Sb)/P, and Nd/La substitutions, the $T_{\rm c}$ enhancement with these substitutions can be roughly interpreted as a result of increase in electron correlation. 
O/F(H) substitution has two effects, namely, carrier doping and increase in $h_{Pn}$. 
The former weakens the electron correlation, while the latter strengthens it. 
The apparent separation of SC1 and SC3 in the La-system (Fig.5(a)) is a result of the competition of these two effects. 
The suppressed $T_{\rm c}$ in this region can be recovered once $d_{xy}$ FS is recovered by Sb-substitution. 
In Fig. 6, we can see that the electron correlation is enhanced towards the optimal point. 
$T$-linear resistivity observed in SC3 (Fig. 3) is a common property of strongly-correlated superconductors. 
Another strength of this model is that the electron correlation depends on the orbital, and the $d_{xy}$ orbital plays a crucial role. 
This is consistent with the present results for SC1 and SC3.

The evidence for strong coupling regime in this superconductor can be seen in the so-called Uemura plot\cite{uemura1991basic}. This famous plot demonstrates that many exotic superconductors, including cuprates, show a linear correlation between $T_{\rm c}$ and muon relaxation rate $\sigma$. 
This suggests that all exotic superconductors are far from BCS superconductors but near the BCS-BEC crossover. 
Shortly after the discovery of Fe-based superconductors, $\sigma$ was determined for several $R$FeAs(O,F) ($R$ = La, Nd and Ce) compounds\cite{carlo2009static}. 
All the data points for the measured compounds lie on the Uemura line, indicating that these superconductors are categorized as strong-coupling superconductors.

On the other hand, $\sigma$ for the low $T_{\rm c}$ ($\sim$ 5 K) compound LaFePO was found to be comparable to that for other high $T_{\rm c}$ ($\sim$ 30 K) compounds\cite{carlo2009static}. 
This means that the LaFePO data point in the $T_{\rm c}$-$\sigma$ plot is far from the Uemura line. 
Recently, we performed systematic measurements of $\sigma$-values for various compounds of LaFeAs$_{1-x}$P$_x$(O,F), and found that with increasing As-content the data point approaches the Uemura-line \cite{miyasaka2020}. 
This implies that the electron correlation increases with increasing As-content, as we expect from this study, and as Misawa $et$ $al.$ predicted in their calculations\cite{misawa2012ab,misawa2014superconductivity}. 
Therefore, in the SC2 phase, the system changes from a weak coupling BCS superconductor to a strong coupling non-BCS phase with increasing As-content, while SC1 and SC3 are the phases in the strong coupling regime, where the $d_{xy}$ hole FS plays a major role.

\section*{Conclusion}

The electronic phase diagram was extensively investigated for $R$FeAs$_{1-x}$(P/Sb)$_x$O$_{1-y}$(F/H)$_y$ ($R$ = La and Nd) over a wide composition range. 
It was observed that the pnictogen height ($h_{Pn}$) increases with increasing As/Sb-composition and electron doping, and that $h_{Pn}$ in the Nd-system is larger than that in the La-system. 
The changes in the superconducting transition temperature ($T_{\rm c}$) with composition can be understood in terms of the contribution of the $d_{xy}$ hole Fermi surface (FS) and electron correlation strength, both of which are controlled by $h_{Pn}$. 
The apparently-separated two superconducting phases (SC1 and SC3) in the La-system merged with Sb-substitution, suggesting that the origins of these two phases are the same in a strong coupling regime. 
While FS nesting controlled by band filling plays an important role in the weak coupling superconductivity for the P-rich compositions (SC2 phase), as $h_{Pn}$ increases with increasing As/Sb-content, superconductivity regime switches to strong coupling one with $d_{xy}$ orbital playing an important role (SC1/SC3), which supports the orbital-selective Mott scenario.

\section*{Methods}
Polycrystalline $R$FeAs$_{1-x}$(P/Sb)$_x$O$_{1-y}$F$_y$ ($R$ = La and Nd) were synthesized using solid state reaction methods. 
For $R$ = La, a stoichiometric-ratio mixture of LaAs, LaSb (or LaP), As (or P), Fe$_2$O$_3$, Fe and LaF$_3$ powder was pressed into a pellet in a pure Ar-filled glove box and heated at 1100-1250 $\tccelsius$ for 40-60 hours in an evacuated silica tube. 
For $R$ = Nd, a mixture of NdAs, NdP, Fe$_2$O$_3$, Fe and FeF$_2$ powder was used for synthesis. 
As the solubility limit of fluorine is low ($y<0.15-0.20$), hydrogen was used to synthesize heavily electron-doped samples ($y=0.20-0.40$). 
H-substituted samples were synthesized under high pressure. 
Among the above powders, LaF$_3$ and FeF$_2$ were replaced by LaH$_2$ and NdH$_2$ for the La- and Nd-systems, respectively. 
A mixture of all the chemicals with the appropriate stoichiometric ratio was pressed into a pellet and heated at 1100 $\tccelsius$ for 2 hours under 4 GPa.

All the samples were characterized using high-resolution X-ray diffraction with beam energy of 11.5 keV and 15 keV at room temperature in BL-8A/8B of Photon Factory in KEK, Japan. 
The lattice constants, pnictogen ($Pn$) height from the Fe-plane ($h_{Pn}$), $Pn$-Fe-$Pn$ bond angle ($\alpha$), and Fe-$Pn$ bond distance were calculated from the experimental data by Rietveld analysis\cite{izumi2007three}.

As, P, and Sb concentrations were determined by energy-dispersive X-ray (EDX) spectroscopy. 
In all samples, the estimated As-, P-, and Sb-concentrations were almost the same as the nominal values. 
We could not determine the F and H concentrations by EDX, because F and H are light elements.

Magnetic susceptibility measurements were performed using a Magnetic Property Measurement System (MPMS), with an applied field of 10 Oe. 
Electrical resistivity was measured using a standard four-probe method. 
Most of the $T_{\rm c}$-values presented here were those determined from zero resistivity.

\bibliography{sample8}

\section*{Acknowledgements}

We thank H. Sagayama and R. Kumai for support with the X-ray diffraction measurements at KEK.

\section*{Author contributions statement}
S.M. proposed the study and played a coordinating role. 
T.K., H.T., T.Y., M.U., A.T., and K.T.L. synthesized the samples and measured them.
T.K., S.M., and S.T. discussed the results, and wrote the manuscript.

\section*{Additional information}

\textbf{Competing interests} The authors declare no competing financial interests.


\begin{figure}[ht]
\centering
\includegraphics[width=\linewidth]{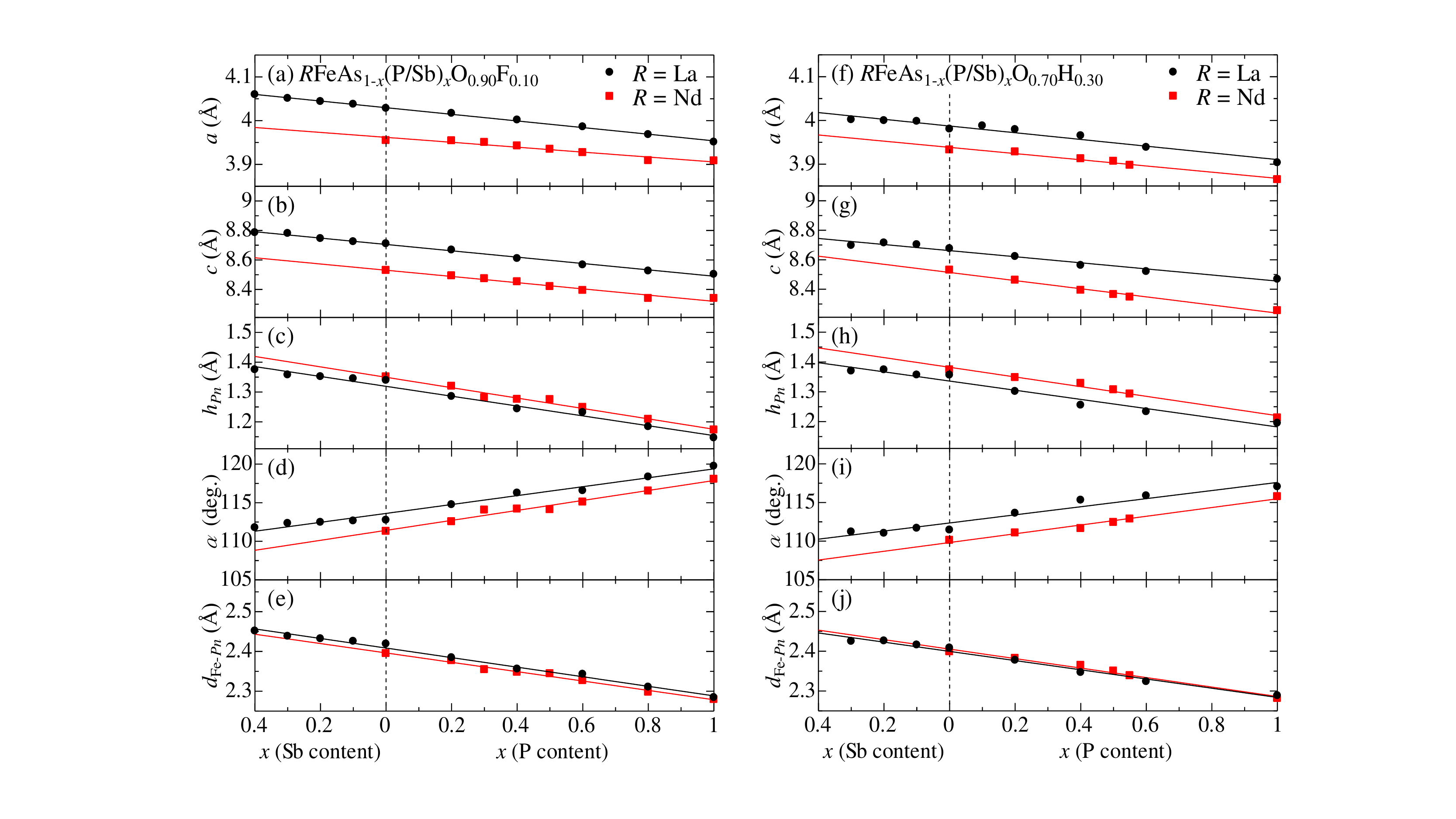}
\caption{
$x$-dependence of structural parameters for (a)-(e) $R$FeAs$_{1-x}$(P/Sb)$_{x}$O$_{0.90}$F$_{0.10}$ and (f)-(j) $R$FeAs$_{1-x}$(P/Sb)$_{x}$O$_{0.70}$H$_{0.30}$ ($R$ = La and Nd). 
(a) and (f) Lattice constant $a$. 
(b) and (g) Lattice constant $c$. 
(c) and (h) Pnictogen height from the Fe plane, $h_{Pn}$. 
(d) and (i) $Pn$-Fe-$Pn$ bond angle $\alpha$. 
(e) and (j) Fe-$Pn$ bond length $d_{{\rm Fe}-Pn}$. 
Structural parameters for other compositions are shown in Supplementary Information (Fig. S1).
}
\label{fig1}
\end{figure}

\begin{figure}[ht]
\centering
\includegraphics[width=\linewidth]{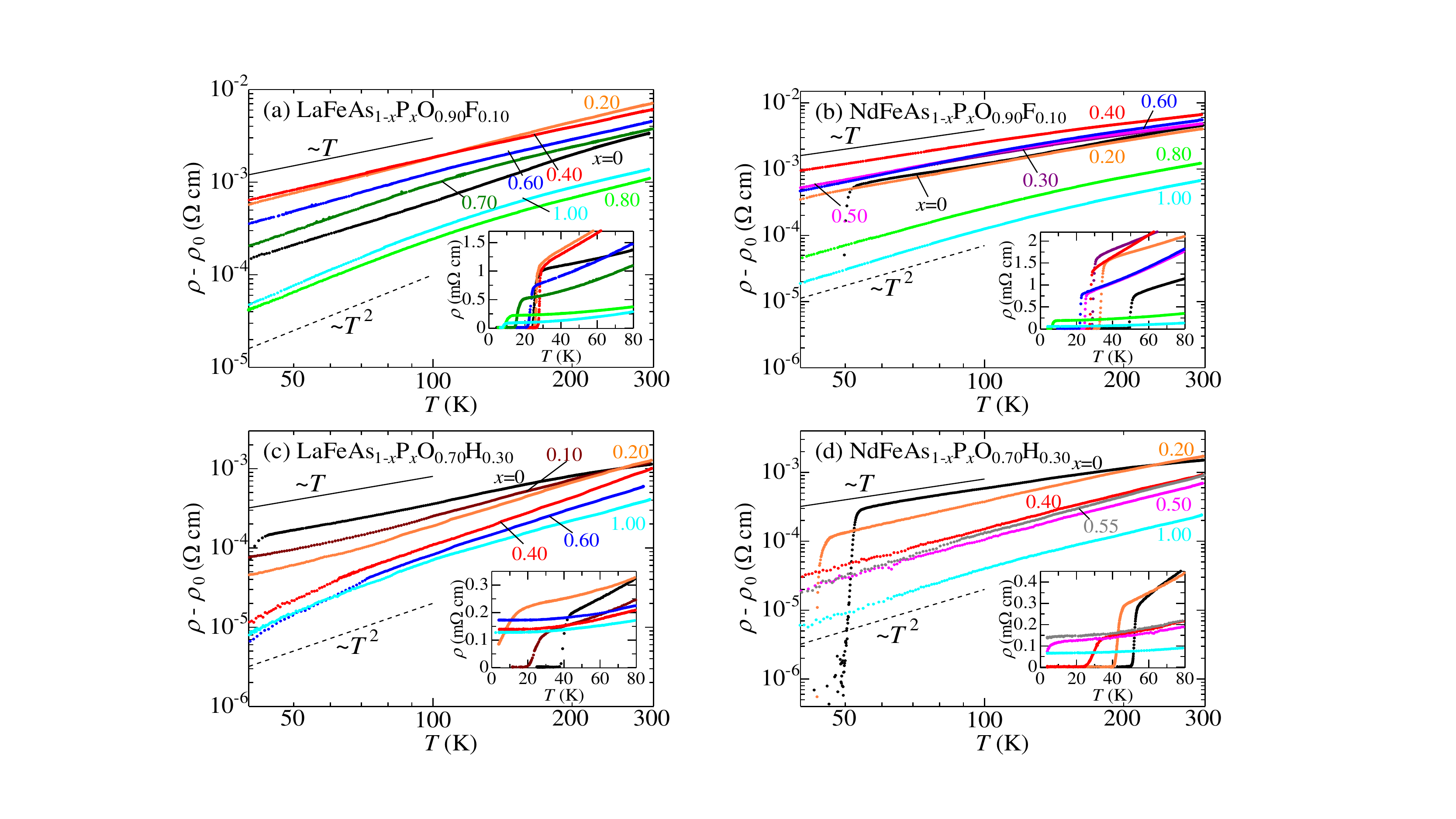}
\caption{
Temperature ($T$-) dependence of $\rho-\rho_0$ for $R$FeAs$_{1-x}$P$_{x}$O$_{0.90}$F$_{0.10}$ and $R$FeAs$_{1-x}$P$_{x}$O$_{0.70}$H$_{0.30}$ ($R$ = La and Nd), where $\rho_0$ is residual resistivity. 
(a) LaFeAs$_{1-x}$P$_x$O$_{0.90}$F$_{0.10}$. 
(b) NdFeAs$_{1-x}$P$_x$O$_{0.90}$F$_{0.10}$. 
(c) LaFeAs$_{1-x}$P$_x$O$_{0.70}$H$_{0.30}$. 
(d) NdFeAs$_{1-x}$P$_x$O$_{0.70}$H$_{0.30}$. 
Insets of (a)-(d) show $T$-dependence of resistivity $\rho$ at low temperatures. 
$T$-dependence of $\rho$ for other compositions are shown in Supplementary Information (Figs. S2 and S3).
}
\label{fig2}
\end{figure}

\begin{figure}[ht]
\centering
\includegraphics[width=\linewidth]{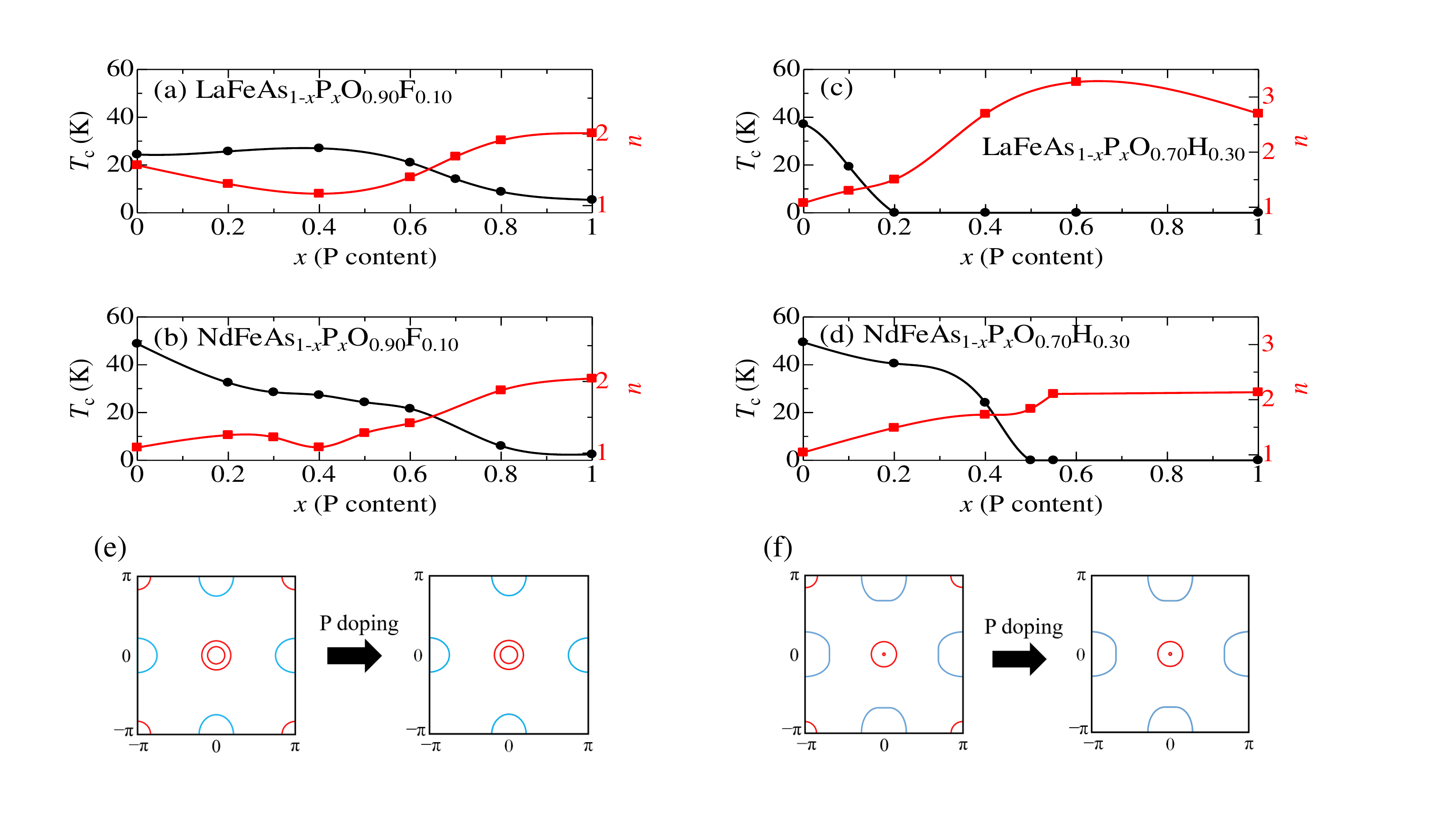}
\caption{
P concentration $x$-dependence of $T_{\rm c}$ (black circles) and $n$ in $\rho (T)=\rho_0 + AT^n$ (red squares), and schematic Fermi surfaces, for $y=$0.1 and 0.3. 
(a) LaFeAs$_{1-x}$P$_x$O$_{0.90}$F$_{0.10}$. 
(b) NdFeAs$_{1-x}$P$_x$O$_{0.90}$F$_{0.10}$. 
(c) LaFeAs$_{1-x}$P$_x$O$_{0.70}$H$_{0.30}$. 
(d) NdFeAs$_{1-x}$P$_x$O$_{0.70}$H$_{0.30}$. 
Solid lines are visual guides. 
(e) and (f) Schematic Fermi surfaces. Solid red and blue lines indicate hole and electron Fermi surfaces, respectively.
}
\label{fig3}
\end{figure}

\begin{figure}[ht]
\centering
\includegraphics[width=\linewidth]{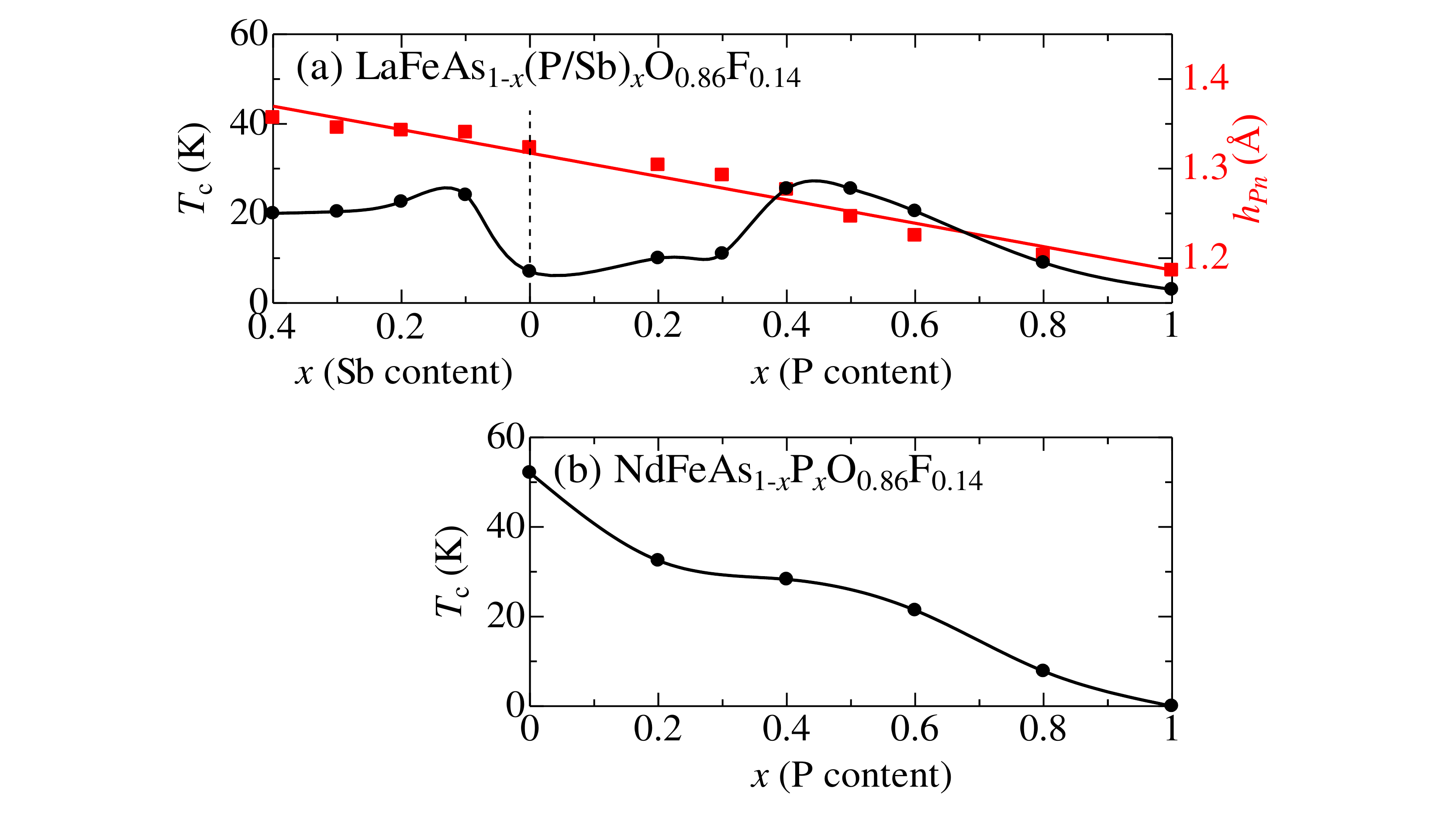}
\caption{
P(Sb) concentration $x$-dependence of $T_{\rm c}$ (black circles) for (a) LaFeAs$_{1-x}$(P/Sb)$_x$O$_{0.86}$F$_{0.14}$ and (b) NdFeAs$_{1-x}$P$_x$O$_{0.86}$F$_{0.14}$. 
The $x$-dependence of $h_{Pn}$ (red squares) for  LaFeAs$_{1-x}$(P/Sb)$_x$O$_{0.86}$F$_{0.14}$ is also shown in the panel (a). 
Solid lines are visual guides.
}
\label{fig4}
\end{figure}

\begin{figure}[ht]
\centering
\includegraphics[width=\linewidth]{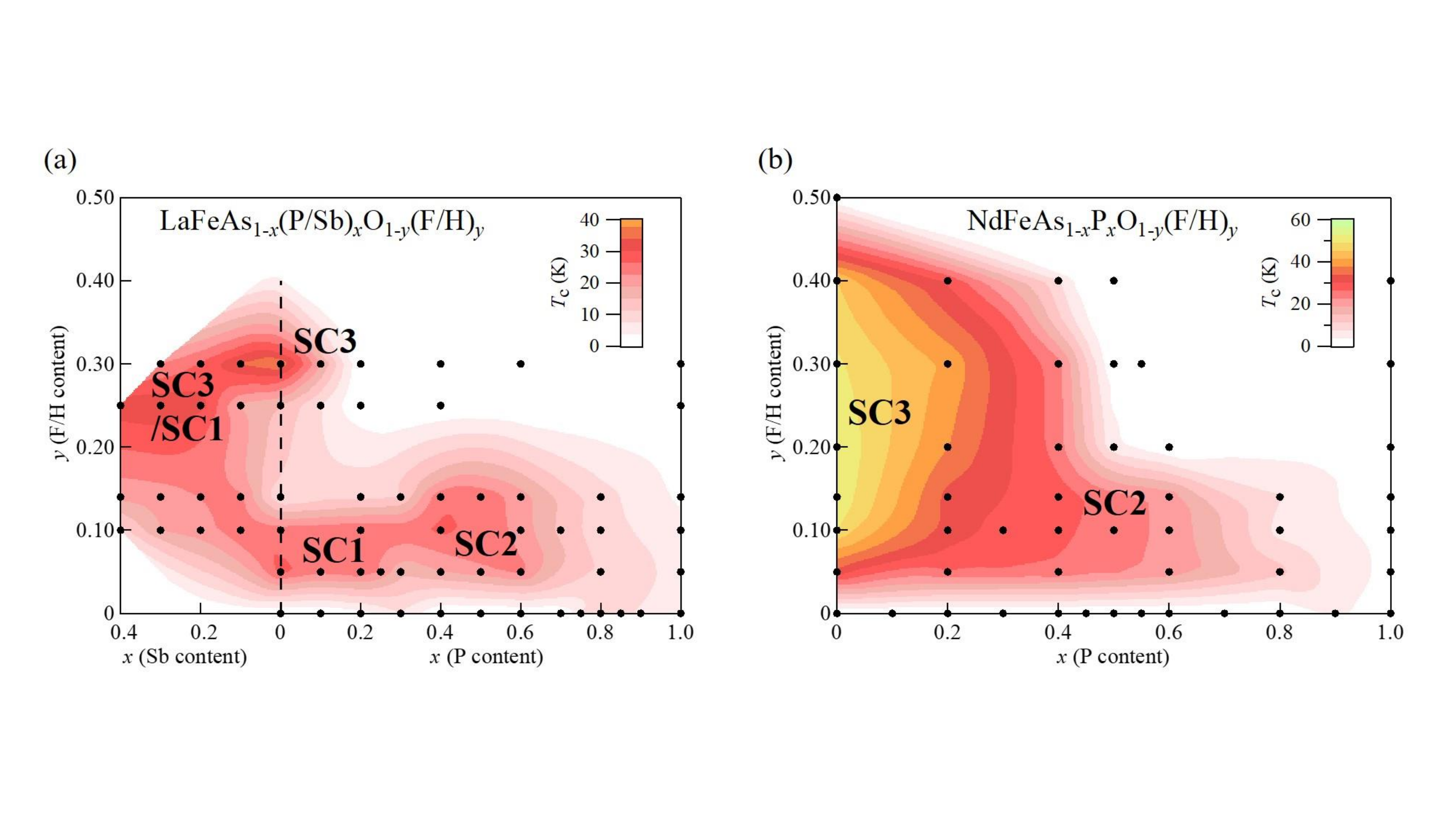}
\caption{
Schematic phase diagrams of (a) LaFeAs$_{1-x}$(P/Sb)$_x$O$_{1-y}$(F/H)$_{y}$ and (b) NdFeAs$_{1-x}$P$_x$O$_{1-y}$(F/H)$_{y}$. 
Dots represent the compositions examined in this study. 
Contour lines of $T_{\rm c}$ are drawn by extrapolation using the data of Figs. 3 and 4, supplementary materials, and previous results\cite{miyasaka2017three,miyasaka2013two,Lai2014evolution}.
}
\label{fig5}
\end{figure}

\begin{figure}[ht]
\centering
\includegraphics[width=\linewidth]{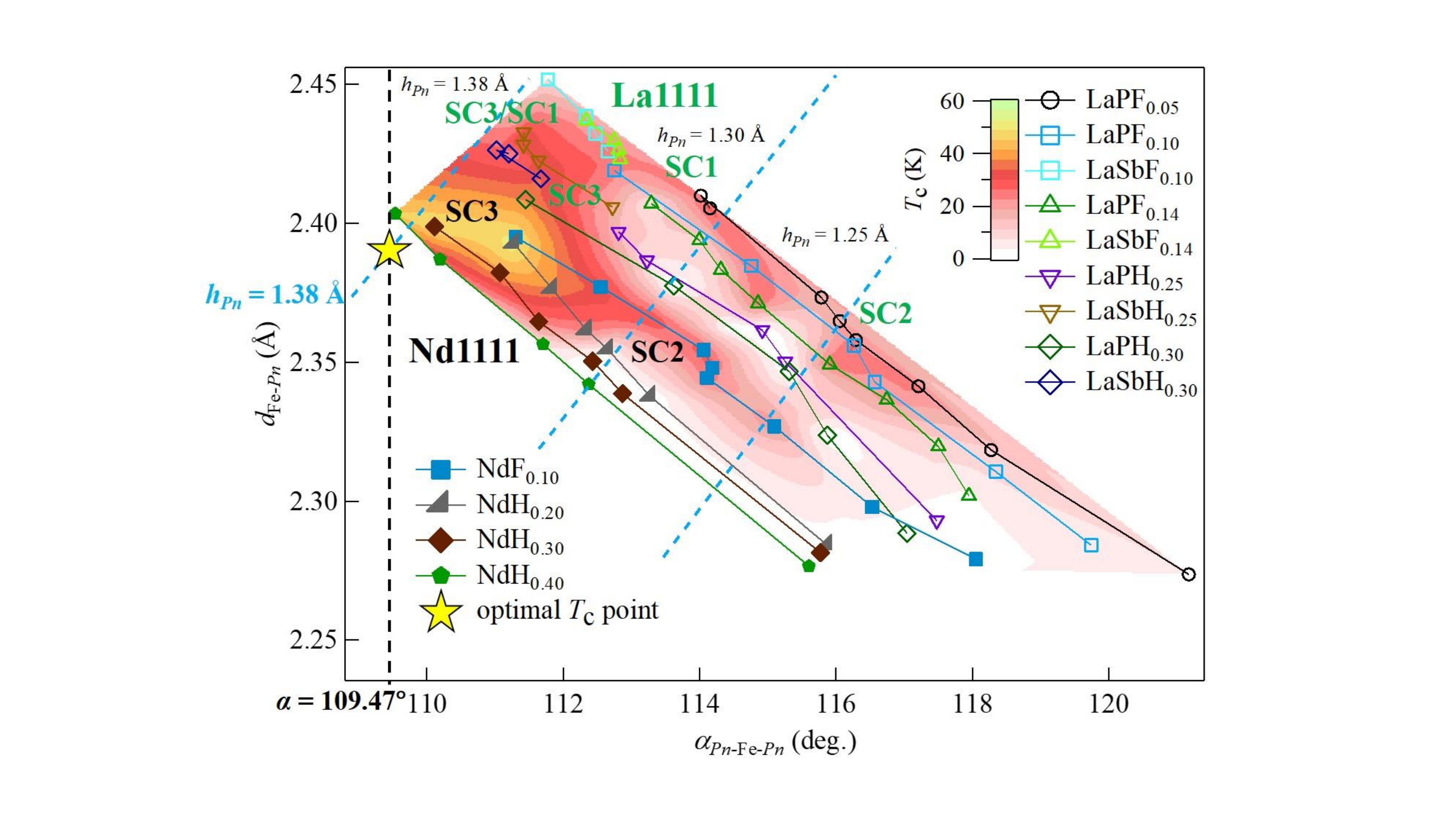}
\caption{
Contour plot of $T_{\rm c}$ for $R$Fe(As,P/Sb)(O,F/H) systems ($R$ = La and Nd). 
Symbols P, Sb, F, and H represent P, Sb, F, and H substitution, respectively. 
Blue broken lines indicate constant values of pnictogen height from the Fe plane, $h_{Pn}=1.25$, 1.30, and 1.38 \AA. 
The black broken line shows the constant value of the $Pn$-Fe-$Pn$ bond angle $\alpha=109.47^\circ$. 
The yellow star indicates the optimal local crystal structure point.
}
\label{fig6}
\end{figure}

\end{document}